\documentclass[12pt]{article}
\usepackage{graphicx}
\usepackage{caption}
\usepackage{amssymb}
\usepackage[numbers,sort&compress,square]{natbib}
\setcitestyle{numbers}
\setcitestyle{square}
\usepackage{url}
\usepackage{color}
\def\vareps{\varepsilon}
\def\be{\begin{equation}}
\def\ee{\end{equation}}
\def\({\left(}
\def\){\right)}
\def\d{{\rm d}}

\usepackage[ddmmyyyy]{datetime}
\def\be{\begin{equation}}
\def\ee{\end{equation}}

\def\({\left(}
\def\){\right)}

\begin{document}
\begin{center}
{\Large Vibrational angular momentum level densities of linear molecules}
\vspace{1cm}

{\large Klavs Hansen\footnote{Corresponding author; email: klavshansen@tju.edu.cn}}\\
\vspace{0.3cm}
Center for Joint Quantum Studies and Department of Physics, 
School of Science, Tianjin University, 
92 Weijin Road, Tianjin 300072, China\\
and\\
Quantum Solid-State Physics, Department of Physics and Astronomy,\\
KU Leuven, 3001 Leuven, Belgium\\
\vspace{1cm}

{\large Piero Ferrari}\\
\vspace{0.3cm}
Quantum Solid-State Physics, Department of Physics and Astronomy,\\
KU Leuven, 3001 Leuven, Belgium\\

\today,~\currenttime
\end{center}

\begin{abstract}
While linear molecules in their vibrational ground state 
cannot carry angular momentum around their symmetry axis, 
the presence of vibrational excitations can induce deformations 
away from linearity and therefore also allow angular momentum 
along the molecular axis.
In this work, a recurrence relation is established for the 
calculation of the vibrational level densities (densities of 
states) of linear molecules, specified with respect to both 
energy and angular momentum.
The relation is applied to the carbon clusters of sizes $n=4,6,7$
as a case study.
\end{abstract}

\section*{Introduction}
\label{intro}
It is a commonly accepted truth that linear molecules can only 
rotate around two axes.
As stated this is true. 
The reason is the simple fact that the principal moment of 
inertia around the molecular symmetry axis is zero, 
or at least as close to zero as the presence of off-axis 
electrons allows. 
Disregarding the rotational motion of electrons, which 
have quantum energies that are usually far beyond molecular 
rotational energies, no molecular rotation can occur around an 
axis on which all nuclei in a molecule are located.
The statement is particularly transparent in quantum theory 
with its quantized angular momentum \cite{Atkins2011}.

This simple result does not, however, imply that a linear 
molecule can only carry angular momentum in the two 
directions perpendicular to the symmetry axis.
Excitation of the doubly degenerate perpendicular vibrational 
modes will induce deformations of the molecule away from 
linearity and generate non-zero moments of inertia, hence
allowing for a non-zero angular momentum around the symmetry 
axis.

This is a special case of the general possibility of 
vibrations carrying angular momentum which was already 
established for the eigenmodes of an elastic 
sphere with the work in ref. \cite{Lamb1881}.
Some of these predicted vibrational modes were only confirmed 
much later \cite{KuokPRL2003}.
Also the presence of angular momentum in the elementary 
excitations of helium droplets play a role for their thermal 
properties, as analyzed in detail in \cite{HansenPRB2007}.

The spectroscopic implications of the phenomenon in molecular
context has been the subject of a number of studies. 
Studies of non-linear molecules were reported in \cite{WilsonJCP1935,
NielsenPR1949,NielsenRMP1951,OkaJCP1967,WatsonMP1968,MillsMS1972,
NemesAPH1984,MielkeJPCA2013,PanJPCA2016}, 
and linear molecules in \cite{NielsenRMP1951,EsplinAO1989,
Alvarez-EstradaPLA1991,HiranoJMS2018},
including experimental studies of both spectroscopic nature 
and fragmentation processes where the vibrational angular momentum 
plays a role \cite{EsplinAO1989,MordauntJCP1998}.
Most of this previous work on the subject has been dedicated to 
the energies and degeneracies of the single 
modes for spectroscopic purposes.
The question has also been treated for bulk matter \cite{McLellanJPC1988}, 
for which phonon interactions with spin is of interest \cite{ZhangPRL2014}.

In contrast, scant attention has been devoted to the effect of the 
vibrational angular momentum of molecules in connection with reactivity
(see \cite{McLellanJMS1989}, though).
However, the energy and angular momentum resolved level densities
are essential in order to implement the relevant conservation laws 
in thermally activated reactions of both bi- and unimolecular
nature.

For a linear molecule composed of $n$ atoms, there are $n-1$ 
stretching modes ($n$ displacements of $n$ atoms along the 
molecular axis, of which one mode is a translational motion) and $2(n-2)$ 
bending modes ($2n$ modes, of which two with no nodes in the 
CM system are translational, and two with one node are 
rotations around the two axes perpendicular to the molecular axis),
in agreement with the standard counting of non-vibrational modes
(the term bending mode or perpendicular mode/motion will be used 
to designate all these $2(n-2)$ modes, although a number of these 
modes are really best described as transverse waves).
The term molecular axis will refer to the ground state 
axis around which the molecules oscillate in the $2(n-2)$
bending modes.

The enumeration places the degrees of freedom of a linear 
molecule in four classes. 
One comprises the three translational degrees of freedom,
which decouple rigorously from all other degrees of freedom 
by the translational invariance of the equations of motion.
A second contains the two rotational motions around the axes 
perpendicular to the molecular axis.
The third class contains the vibrational motion along the 
molecular axis (longitudinal motion, $n-1$ modes).
These vibrational modes do not carry any angular momentum.
The fourth class comprises the $2(n-2)$ bending modes
of interest here, carrying both vibrational energy and 
angular momentum.

The level density, which is the focus of this article, counts 
the number of states for a given energy and is, therefore, the 
microcanonical partition function.
For small, isolated systems it assumes a particularly important 
role because the difference between canonical and microcanonical 
quantities become important for small systems. 
This holds for issues concerning the energy content but 
perhaps even more for questions concerning angular momentum,
as this quantity does not appear in the description of the 
macrostate of a canonical system.

The total angular momentum resolved vibrational level 
density, $\rho_{tot}(E,L_z)$, is obtained as a convolution 
of the function for the longitudinal modes, $\rho_l(E)$,
and the ones for the bending motion, $\rho_p(E,L_z)$,
as $\rho_{tot}(E,L_z)=\int \rho_l(E-\epsilon) 
\rho_p(E,L_z) \d \epsilon$.
An implementable procedure for this will be given below,
and its use demonstrated with a calculation of three small
carbon clusters.

\section*{Angular momentum along the z-axis, $L_z$}

The angular momentum eigenstates for a bending mode in the 
harmonic approximation, which will be considered sufficient here, 
are those of a two-dimensional harmonic oscillator due to the 
double degeneracy of the bending modes.
It is an interesting fact that the large quantum number 
limits of such systems are not the classical limits, and that 
semiclassical quantization does not give the correct answers 
in that limit \cite{LiuCTP2000}.
As it is, the exact spectrum can be found in the harmonic 
approximation without taking this limit.
To do so, each degenerate pair of bending modes can be 
considered separately.
Orienting the coordinate system with the $z$-axis along the 
molecular axis, the operator for the $z$-projection 
of the angular momentum is written as
\begin{eqnarray}
L_z &=& xp_y -yp_x \\\nonumber
&=& \frac{{\rm i} \hbar}{2} \(a_x^{\dagger} + a_x \)
\(a_y^{\dagger} - a_y\) 
- \frac{{\rm i} \hbar}{2} \(a_y^{\dagger} + a_y \)
\(a_x^{\dagger} - a_x\)\\\nonumber
&=& {\rm i} \hbar \( a_y^{\dagger} a_x - a_x^{\dagger} a_y \),
\end{eqnarray}
where $a_x^{\dagger},a_x$ are the raising and lowering 
operators for the oscillator vibrating along the $x$-direction, 
and similarly for the operators with subscript $y$.
$L_z$ commutes with the Hamiltonian:
\be
[L_z, H] = {\rm i} \hbar^2\omega [\( a_y^{\dagger} a_x - 
a_x^{\dagger} a_y \),
a_y^{\dagger} a_y + a_x^{\dagger} a_x +1] = 0.
\ee
The Hilbert space of the transverse modes is spanned by the 
vibrational states $|n,m \rangle$, where the first integer 
gives the energy of the motion in the $x$-direction and the 
second in the $y$-direction.
It is clear from the expression for $L_z$ that $|n,m\rangle$ 
states are not eigenstates of $L_z$. 
An explicit calculation shows this:
\begin{eqnarray}
L_z |n,m\rangle &=& {\rm i}\hbar \( a_y^{\dagger} a_x - 
a_x^{\dagger} a_y \) |n,m \rangle 
\\\nonumber
&=& {\rm i}\hbar \(\sqrt{m+1}\sqrt{n}|n-1,m+1 \rangle - 
\sqrt{n+1}\sqrt{m} |n+1,m-1 \rangle \).
\end{eqnarray}
As the operator raises one vibrational quantum 
number and lowers the other, and the two oscillators are degenerate,
it is equally clear that the angular momentum eigenstates are 
composed of eigenstates with the same energy, $n+m =$ constant,
of which there are $n+m+1$. 
A direct diagonalization of the three lowest levels with energies 
1, 2 and 3 times $\hbar \omega$ (zero point energy included) 
gives the angular momenta 
0, $\pm \hbar$, and $0,\pm 2\hbar$. 
The general case can be calculated by construction of the operators 
\cite{LiuPRA1998,MotaJPA2002}
\begin{eqnarray}
a_d \equiv \frac{1}{\sqrt{2}}\(a_x - {\rm i} a_y \),
\\\nonumber
a_g \equiv \frac{1}{\sqrt{2}}\(a_x + {\rm i} a_y \),
\end{eqnarray}
and the associated number operators
\begin{eqnarray}
n_d \equiv a_d^{\dagger}a_d,\\\nonumber
n_g \equiv a_g^{\dagger}a_g.
\end{eqnarray}
The Hamiltonian and the angular momentum operators can be written
in terms of these as
\begin{eqnarray}
H = \hbar \omega \( n_d +n_g +1\), \\\nonumber
L_z = \hbar \(n_d - n_g \).
\end{eqnarray}
For a given energy, the angular momentum states therefore differ 
by $2\hbar$, as the calculated example also suggested. 
This spacing, together with the number of states, defines the 
angular momentum spectrum uniquely as having the eigenvalues
$N\hbar, (N-2)\hbar, ....,-N\hbar$ for states 
with the total energy $E = \hbar \omega \(N+1\)$.

\section*{Level densities}

For molecules or clusters for which the vibrational spectrum 
is known, the vibrational contribution to the level density, 
$\rho$, can be calculated with the Beyer-Swinehart algorithm 
\cite{BeyerCACM1973}. 
The algorithm is a convolution of level densities of the 
individual uncoupled degrees of freedom in the form of 
normal modes. 
Due to the discrete nature of vibrational excitations,
this convolution takes the form of a sum over levels. 
The summation is repeated recursively for each new mode
included.
Energy is the only physical argument in the procedure.
The intermediate steps in the recurrence are labeled by the 
number of modes, $M$, that have been included. 
This adds the integer $M$ as an argument to the function 
\cite{BeyerCACM1973}:
\be
\rho = \rho(M,E).
\ee
The procedure can be condensed by contracting terms on the 
right hand side and the resulting recurrence written 
as \cite{HansenJCP2008}
\be
\label{BS}
\rho(M,E) = \rho(M,E-\hbar\omega_M)+\rho(M-1,E).
\ee
Adding angular momentum expands this array with that quantum 
number.
In total, the procedure then makes use of the energy, $E$, 
the angular momentum $L_z$, and the integer $M$ labeling 
the vibrational modes included in the recurrence:
\be
\rho = \rho(M,E,L_z).
\ee
As usual, angular momentum is most conveniently given in 
units of $\hbar$ and is therefore represented by integers.
This will be used for the argument of $\rho$.
For this choice, the dimension of $\rho$ is energy to the 
power -1, where the unit of energy is determined by the 
value chosen for the input vibrational frequencies. 
For other purposes than indexing $\rho$, the dimension of 
$L_z$ will be that of Planck's constant.

The rotating bending modes are included by extending summations 
over modes to include all possible angular momenta. 
As the modes come pairwise, the mode index $M$ changes by 
two for each iteration. 
Adding one more quantum of energy will change the angular 
momentum by either +1 or -1:
\begin{eqnarray}
\rho(M,E,L_z) &=& \rho(M-2,E,L_z)\\ \nonumber
&+&\sum_{i=0}^{1} \rho(M-2,E-\hbar \omega_M,L_z+2i-1)\\\nonumber
&+&\sum_{i=-1}^{1} \rho(M-2,E-2\hbar\omega_M,L_z+2i)\\\nonumber
&+&\sum_{i=-1}^{2} \rho(M-2,E-3\hbar \omega_M,L_z+2i-1)\\\nonumber
&+&\sum_{i=-2}^{2} \rho(M-2,E-4\hbar\omega_M,L_z+2i)\\\nonumber
...
\end{eqnarray}
The first term on the right hand side of this equation 
is the contribution from the
partitionings where there is no excitation in either of the two 
added modes.
The second term gives the two contributions from a single 
excitation in the two modes, distributed as a negative or a 
positive angular momentum quantum. 
The following terms are generated after the same principle, 
with $n+1$ different contributions to the angular momentum 
for an added energy of $n$.

Also this recurrence relation can be condensed.
The sum contains a subset of terms that adds up to 
$\rho(M,E-\hbar\omega_M,L_z-1)$, and one that adds up to 
$\rho(M,E-\hbar\omega_M,L_z+1)$.
Collecting these two terms leaves out the first term on the right 
hand side, $\rho(M-2,E,L_z)$, and it double-counts terms that on 
inspection are found to add up to $\rho(M,E-2\hbar\omega_M,L_z)$. 
By adding the missing and subtracting the double-counted terms, 
it is therefore possible to write the recurrence in the much 
more compact form
\begin{eqnarray}
\label{Lzfinal}
\rho(M,E,L_z) &=&
\rho(M-2,E,L_z)+ \rho(M,E-\hbar\omega_M,L_z-1)\\\nonumber
&+&\rho(M,E-\hbar\omega_M,L_z+1)
-\rho(M,E-2\hbar\omega_M,L_z).
\end{eqnarray}
A numerical calculation with this expression comprises three 
loops nested in the order of appearance of the arguments in 
$\rho$.
Computationally, two matrices with indices $E,L_z$ are needed. 
If only the modes carrying angular momenta are required, 
the recurrence is started with the initial values 
$\rho(0,j,k) = \delta_{j,0}\delta_{k,0}$, where the $\delta$'s
are Kronecker's $\delta$.
If the non-angular momentum-carrying modes should be included 
into the level density, Eq.\ref{BS} is used with the 
angular momentum index set to zero.
After convoluting the $M$ non-angular momentum modes the result, 
$\rho_l(M,E,0)$, is used as the initial conditions for the 
recurrence in Eq.\ref{Lzfinal} instead 
of $\rho(0,0,0)=1$ etc., i.e. changing the initial conditions to
$\rho(0,E,j) = \rho_l(M,E,j)\delta_{j,0}$. 

\section*{Application to C$_{4,6,7}$}

We will use three small carbon clusters as case studies, 
C$_{4,6,7}$, as representative examples of linear carbon 
clusters.
For the neutral species, linear structures are the lowest 
energy carbon cluster conformers up to $n=7$ \cite{yen2015},
and all three clusters are expected to have linear ground states, 
without any Jahn-Teller deformations.

The ground state and the vibrational frequencies of the 
clusters C$_n$ ($n=4,6,7$) 
were calculated with the ORCA 4.2 software package 
\cite{ORCA2012}. 
The B3LYP exchange-correlation functional was employed, 
in conjunction with the 6-311G* basis set. 
This functional has been successfully used in calculations 
of small carbon clusters, both neutrals and cations 
\cite{martinez2018}.
As shown in previous studies, clusters with an even (odd) 
number of atoms have a triplet (singlet) electronic ground-state 
\cite{Saykally1998}.
We have assumed those multiplicities for the calculations. 

\begin{figure}[ht]
\includegraphics[width=0.5\textwidth]{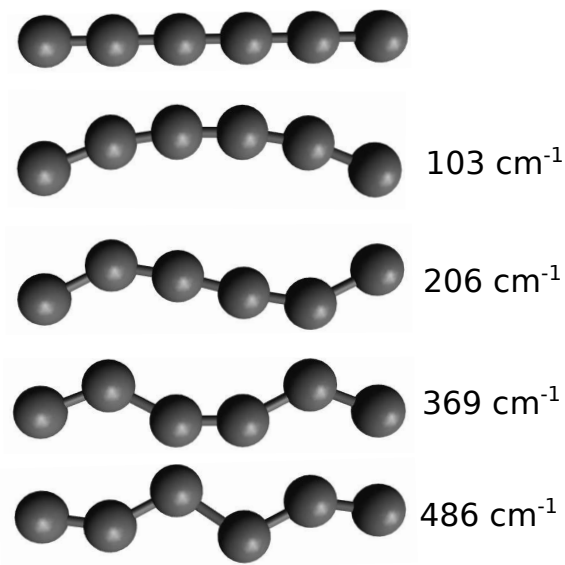}
\begin{centering}
\caption{\label{c6modeexample}The relative amplitudes of the 
normal coordinates for the four doubly degenerate vibrational 
modes of C$_6$. 
The representations of the modes are chosen so all motion of 
the atoms is in one plane.
The top geometry corresponds to the vibrational ground-state 
structure.}
\end{centering}
\end{figure}

Fig.(\ref{c6modeexample}) shows the relative amplitudes 
of the atomic displacement for the four pairs of 
degenerate bending modes of C$_6$, together with the 
energy of the mode (in cm$^{-1}$), calculated for the 
electronic ground state. 
At the top of the figure the vibrational ground-state 
structure is shown, with aligned and almost equidistant 
carbon atoms.

The calculated vibrational frequencies are listed in 
Table \ref{Freq}, together with the experimental values,
given in \cite{Do2015}. 
While the lowest-energy modes are well reproduced by the 
calculations, some deviations are seen for the higher 
frequency modes. 
Overall, however, the calculated values are in fair agreement 
with the experimental values. 
Many of the modes have a dipole moment of zero, preventing 
detection by infrared spectroscopy. 
For this reason, the analysis here will use the 
calculated frequencies.      

\begin{table}[h!]
  \begin{center}
    \caption{Calculated vibrational quantum energies of C$_4$, C$_6$ 
		and C$_7$ in cm$^{-1}$. 
The doubly degenerate bending modes are indicated in bold. 
		The experimental values \cite{Do2015} are shown below the 
		theoretical values calculated here.}
    \label{Freq}
    \begin{tabular}{ll} 
      n & $\hbar \omega$ (cm$^{-1}$)\\
      \hline
      4 & \textbf{169} \textbf{346} 937 1589 2120\\
      4 Exp. & 160, 339, 1549, 2032\\      
      6 & \textbf{103} \textbf{206} \textbf{369} \textbf{486} 668 1225 1729 2028 2179\\
      6 Exp. & 90, 246, 637, 1197, 1694, 1960, 2061\\
      7 & \textbf{77} \textbf{168} \textbf{264} \textbf{527} 585 \textbf{676} 1112 1600 1980 2212 2245\\
      7 Exp. & 496, 548, 1893, 2128\\
    \end{tabular}
  \end{center}
\end{table}

\begin{figure}[h]
\includegraphics[width=0.6\textwidth,angle=0]{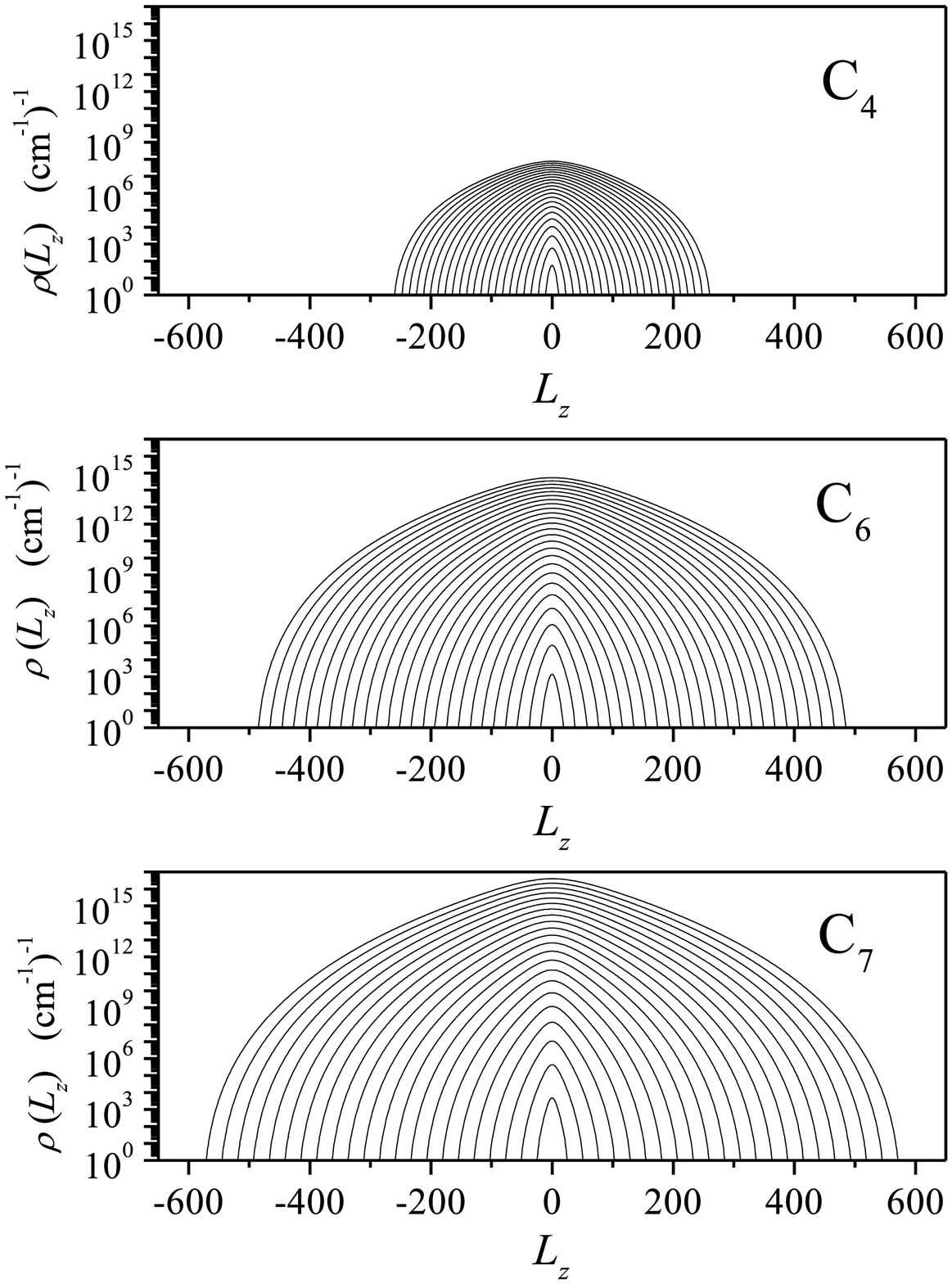}
\begin{centering}
\vspace{-2cm}
\caption{\label{C467Lzld}The angular momenta specified 
level density for different total excitation energies for 
the clusters indicated in the frames, all on identical scales. 
The curves are from top to bottom for $E=44 000$ cm$^{-1}$ 
down to 2000 cm$^{-1}$ in steps of 2000 cm$^{-1}$ in all 
three cases.
$L_z$ is given in units of $\hbar$.}
\end{centering}
\end{figure}

\begin{figure}[h]
\includegraphics[width=0.6\textwidth,angle=270]{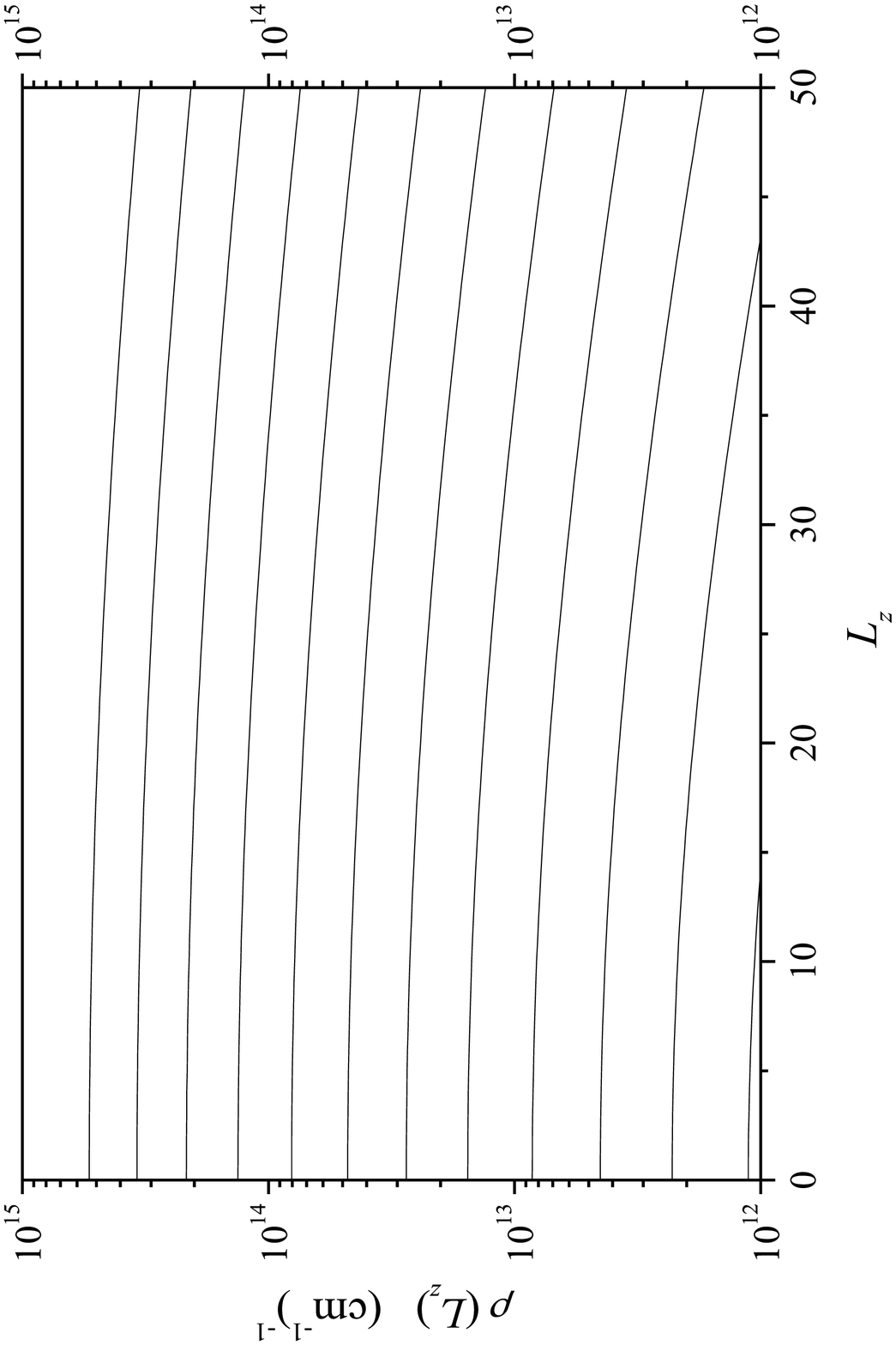}
\begin{centering}
\vspace{-0.5cm}
\caption{\label{C6Lzldzoom} A zoomed view of Fig.(\ref{C467Lzld}) for 
C$_6$.}
\end{centering}
\end{figure}

Fig.(\ref{C467Lzld}) shows the results 
of a calculation of the angular momentum specified level 
densities of C$_{4,6,7}$ with Eq.\ref{Lzfinal} for a series 
of total excitation energies. 
The highest energies in the calculations were chosen to be 
on the order of magnitude of the excitation energies of 
clusters that decay on typical mass spectroscopic time 
scales of tens to hundreds of microseconds.
A zoomed view of the values for C$_6$ around $L_z=0$ is shown
in Fig.(\ref{C6Lzldzoom}).

\section*{Microcanonical temperature, $L_z$-resolved}

Although the values shown in Figs.(\ref{C467Lzld},\ref{C6Lzldzoom}) 
are microcanonical, it is nevertheless possible and 
occasionally also convenient to define a temperature for 
such systems. 
This temperature is particularly useful for the interpretation 
of measured kinetic energy release distributions (KERDs),
because it allows these distributions to be written with 
what is effectively a Boltzmann factor multiplied by phase 
space factors etc. \cite{Hansen2018}.
The definition is (with $k_B$ set to unity)
\be
\label{TMic}
T^{-1} \equiv \left.\frac{\partial 
\rho(E,L_z)}{\partial E}\right|_{L_z}.
\ee

\begin{figure}[ht]
\includegraphics[width=0.6\textwidth,angle=270]{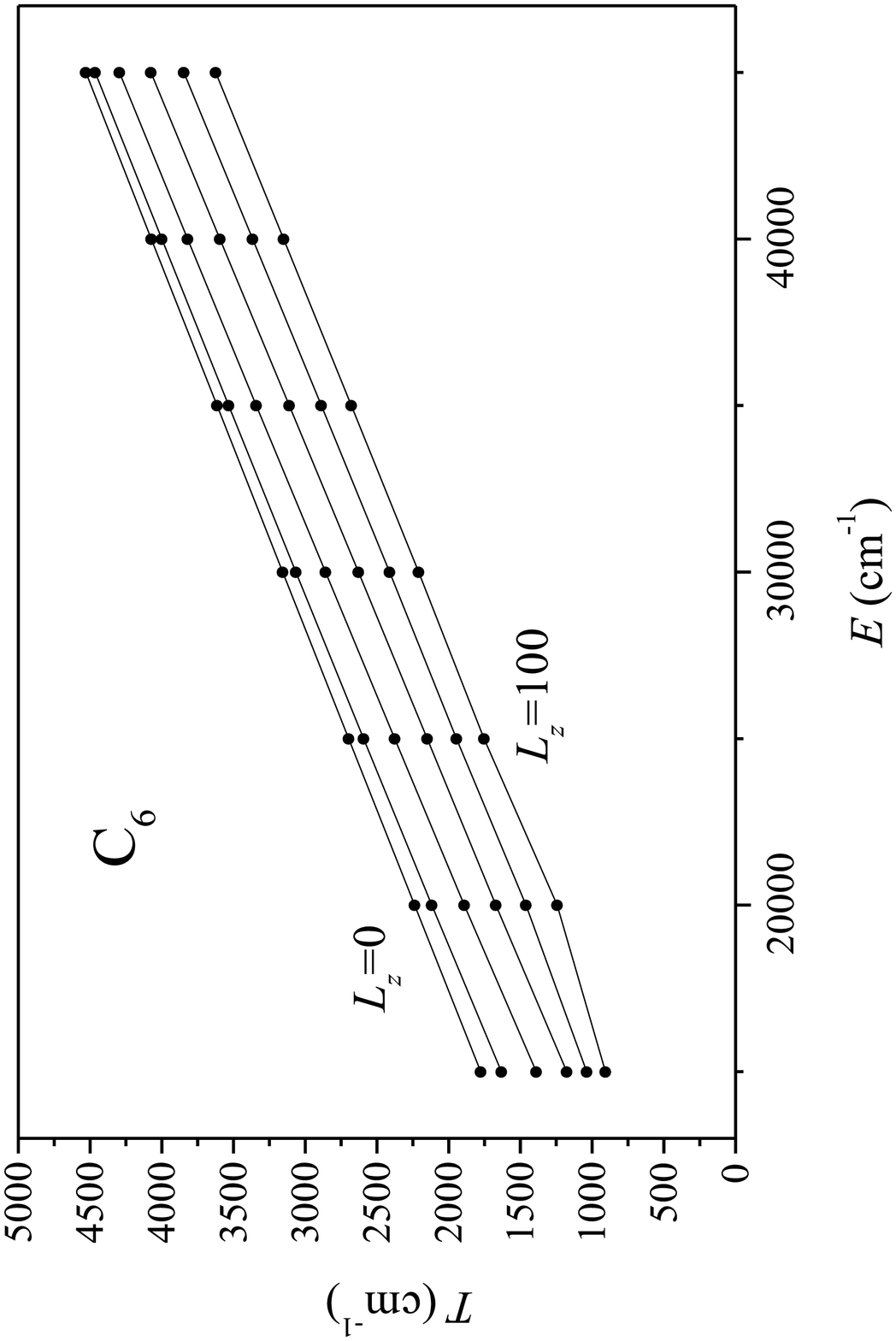}
\begin{centering}
%\vspace{-2cm}
\caption{\label{C6T} The microcanonical temperature for the 
angular momentum specified states.
The values are calculated numerically with a finite derivative 
over 1000 cm$^{-1}$, centered at the abscissa values, with 
Eq.\ref{TMic}.
The curves show the values for $L_z=0 \hbar$ to $100\hbar$, in 
steps of $20\hbar$ (top to bottom).}
\end{centering}
\end{figure}

Fig.(\ref{C6T}) shows the microcanonical temperature for 
different angular momentum sectors of the level density. 
The slopes are clearly all very similar. 
They are also very close to the value of $n-1$ which is expected 
for $n$ harmonic oscillators (the -1 is the difference between 
canonical and microcanonical heat capacities, see
\cite{AndersenJCP2001} for details).
The two other cluster sizes calculated give similar results. 
Although heat capacities are thus fairly insensitive to the 
precise value of the angular momentum, the temperature has a 
clear angular momentum dependence.
The higher the angular momentum, the lower the temperature. 
In conjunction with the constant heat capacity, this 
suggests that the angular momentum comes with a price.
Effectively, it ties up energy and reduces the available 
phase space for the other vibrations. 
For the C$_6$ clusters, the lowering of the 
microcanonical temperature corresponds to a shift in 
energy of about 250 cm$^{-1}$ for each time two units 
of $\hbar$ are added, in the high energy limit. 
This is reasonably close to the average frequency 
of 291 cm$^{-1}$ of the bending modes to 
assign this shift to the excitation required 
to reach the specified angular momentum.

As one application of the results derived, the 
effective Boltzmann factor that enters the kinetic energy 
release distributions will be calculated.
A number of theories are available from the literature for
the calculation of the rest of these distributions. 
A discussion of these theories will lead us too far astray
and we will simply consider the Boltzmann factor here.

We will limit ourselves to consider the expression of a 
product arising from a reactant with zero initial 
vibrational angular momentum.
The function to be approximated is then 
$\rho(E-\vareps,L_z)$, where $E$ is the product energy before 
deduction of $\vareps$, the kinetic energy released, and 
$L_z$ is the resulting vibrational mode angular momentum.
The logarithm of $\rho$ is expanded to first order in 
$\vareps$ to give
\be
\label{KERD1}
\rho(E-\vareps,L_z) \approx 
\rho(E,L_z)\exp\(-\frac{\vareps}{T(E,L_z)}\).
\ee
The temperatures are given by the calculated 
values in Fig.(\ref{C6T}) for C$_6$. 
The temperatures decrease approximately with 11 cm$^{-1}$ 
each time $L_z$ increases by one unit of $\hbar$. 
The approximation sign in Eq.\ref{KERD1} refers to the 
fact that the rate of decrease with $L_z$ for low values 
(below $20\hbar$) is less.
As the values taken by $L_z$ cover a wider range (see top
line of Fig.(\ref{C6Lzldzoom}), we will ignore this and 
represent the temperature for the case of C$_6$ as:
\be
\label{TMiC}
T(E,L_z) = T(E,L_z=0) - 11 \frac{{\rm cm}^{-1}}{\hbar} L_z.
\ee
The coefficient of 11 cm$^{-1}$ is on the order of the 
price in energy per unit of 
angular momentum, which we saw above to be about 250/2 
cm$^{-1}$, divided by the number of vibrational degrees 
of freedom, $3n-5=13$, or 10 cm$^{-1}$, in good agreement 
with the fitted value of 11 cm$^{-1}$.
For the C$_6$ case we can therefore write Eq.\ref{KERD1} 
as
\be
\label{KERD2}
\rho(E-\vareps,L_z) \approx 
\rho(E,L_z)
\exp\(-\frac{\vareps}{T(E,0)-11\frac{{\rm cm}^{-1}}{\hbar} L_z}\).
\ee

The result in Eq.\ref{KERD2} shows that the effective 
temperature as measured with KERDs for a given total 
excitation energy decreases with increasing angular 
momentum.
The $L_z$ dependence at high excitation energies can be 
fitted with the function
\be
\rho(E,L_z) \approx \rho(E,0)\exp\(-a |L_z|^{1.7}\),
\ee
around the peak at $L_z=0$. 
The value of the coefficient is $a = 6\times 10^{-4}$ for the 
highest energy curve shown in Fig.(\ref{C6Lzldzoom}).
This can then be used as is or expanded further with 
$L_z \rightarrow L_z + \delta L_z$.

The factor $\rho(E,L_z)$ is part of the normalization of 
the KERD's, and the entire distribution will then be determined 
by the two exponentials and the prefactor which will not 
be discussed in detail here.
Disregarding this prefactor, the decay will be biased toward a 
reduction of the absolute value of the angular momentum.

\begin{figure}[ht]
\includegraphics[width=0.6\textwidth,angle=270]{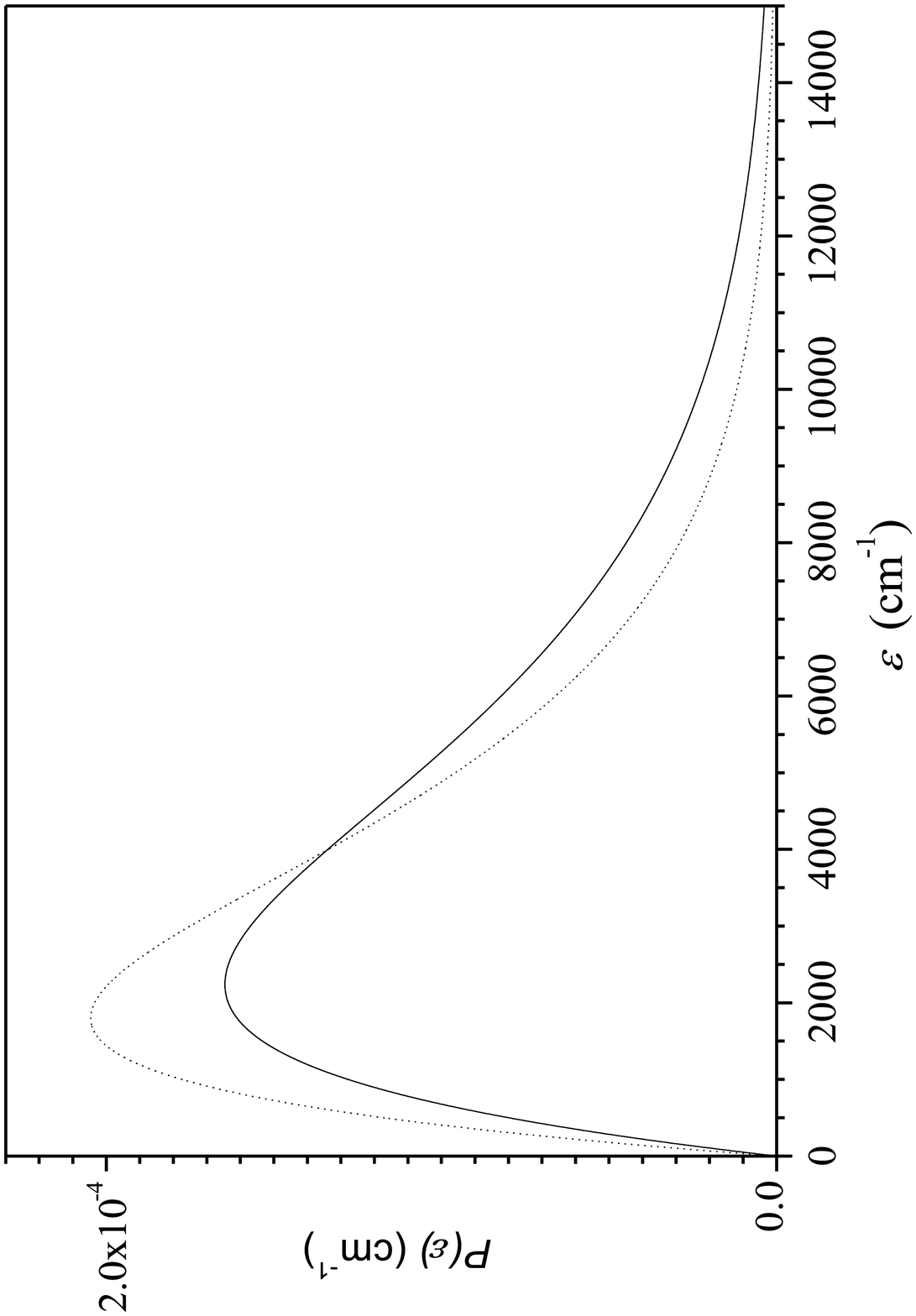}
\begin{centering}
\vspace{-0.8cm}
\caption{\label{KERD} The schematic kinetic energy release 
distributions calculated for C$_6$ with final state energy $E$
and the two $L_z/\hbar$ values of 0 (full line) and 40 (dotted line). 
The energy is 20000 cm$^{-1}$.
The microcanonical temperature is calculated to 2238 cm$^{-1}$
for this energy.
The values of $\rho(E,0)$ and $\rho(E,40)$ differ by a 
factor of ca. 3 for this energy.
The temperatures differ by 20 \%.
The curves are normalized to identical areas.}
\end{centering}
\end{figure}

The $L_z$ distributions will vary with the experimental 
conditions. 
To nevertheless give some information on the effect of the 
vibrational angular momentum, the effect will be 
illustrated with a schematic calculation.
The KER distributions will be represented by the derived 
Boltzmann factor and the preexponential phase space factor 
combined with the kinematic speed factor \cite{Hansen2018}. 
These two factors combine to a factor of kinetic energy to 
the power one.
For the product C$_6$ with the final state energy $E$, 
$L_z$ and fragment translational kinetic energy $\vareps$ 
this becomes:
\be
P(E,L_z,\epsilon) \d \vareps propto \rho\(E,L_z\)\vareps
\exp\( -\frac{\vareps}{T(E,0)- 11\frac{\rm cm^{-1}}{\hbar} L_z}\)
\d \vareps,
\ee
where $L_z$ is the final state vibrational angular momentum.
This is then the expression for zero initial $L_z$.
The curves for $L_z=0, 40~\hbar$, $E=20000$ cm$^{-1}$ are 
shown in Fig.(\ref{KERD}).

\section*{Canonical values}

\begin{figure}[ht]
\includegraphics[width=0.6\textwidth,angle=270]{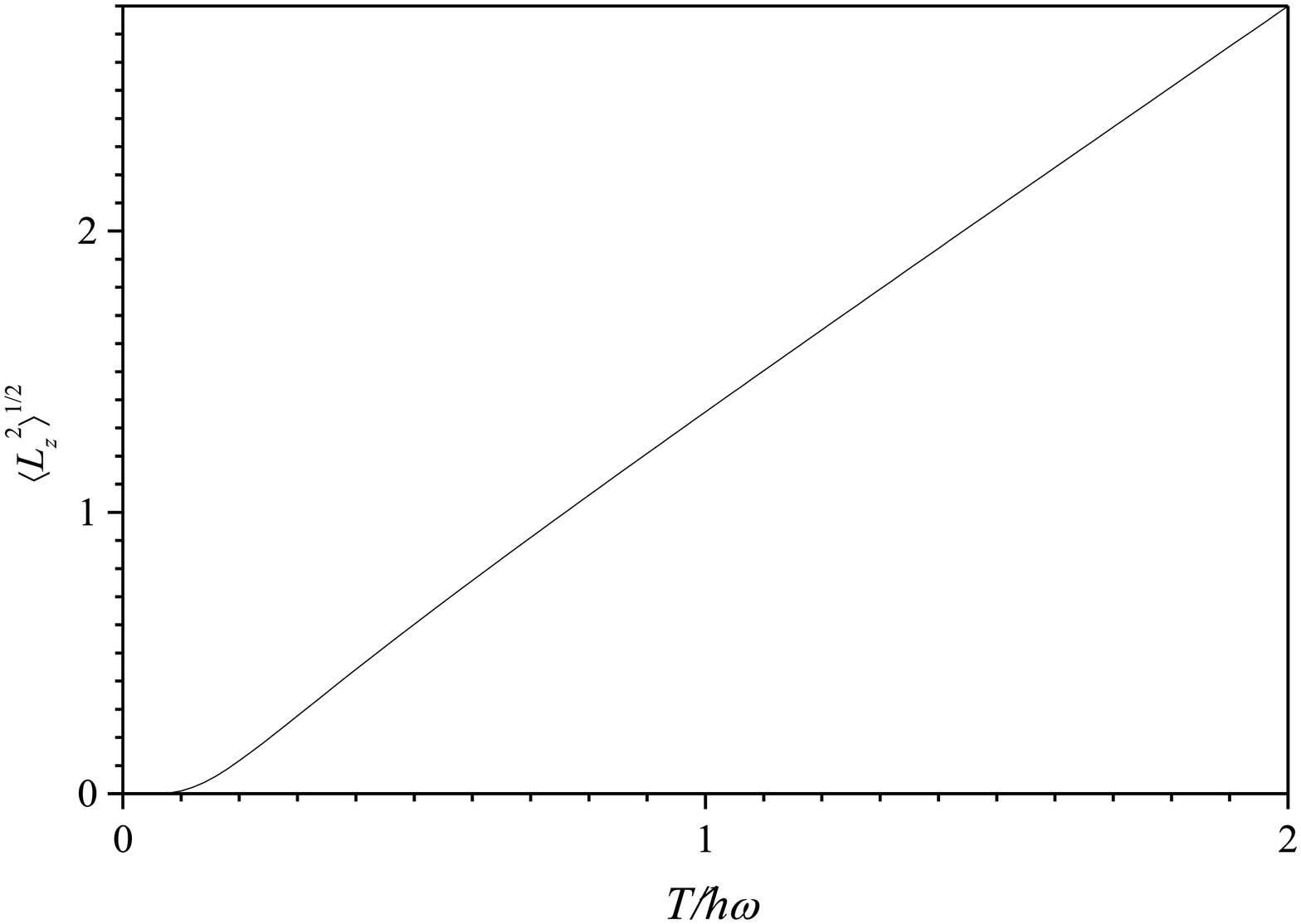}
\begin{centering}
%\vspace{-2cm}
\caption{\label{C6LzRMS} The root-mean-square of $L_z$ in 
units of $\hbar$ as a function of the reduced temperature 
for a single pair of degenerate oscillators.}
\end{centering}
\end{figure}

The canonical equilibrium values are likewise interesting. 
With the level structure found above, the canonical 
partition function is easily calculated and the angular 
momentum specified populations, $P(L_z=n\hbar)\equiv P(n)$ 
found for one degenerate pair of modes with frequency 
$\omega$ to be
\begin{eqnarray}
P(0) &=& \(1-{\rm e}^{-2 \beta \hbar \omega}\)^{-1}\frac{1}{Z},\\
\nonumber
P(\pm n, n>0) &=& {\rm e}^{-n\beta\hbar\omega}P(0),
\end{eqnarray}
where the total partition function is
\be
Z = \(1-{\rm e}^{-\beta\hbar\omega }\)^{-2}.
\ee
The zero of the energy is set equal to the zero point 
energy of the oscillators here.
The mean value of $L_z$ given by these distributions is 
obviously zero. 
The mean square is calculated to 
\begin{eqnarray}
\frac{1}{\hbar^2}\langle L_z^2 \rangle &=& 
P(0) \sum_{n=-\infty}^{\infty} n^2 
{\rm e}^{-|n|\beta \hbar \omega}\\\nonumber
&=& 2P(0) \(\frac{1}{\hbar\omega}\)^2 
\frac{\partial^2}{\partial \beta^2}
\(1-{\rm e}^{-\beta\hbar\omega}\)^{-1}\\\nonumber
&=& 2 {\rm e}^{-\beta\hbar\omega} 
\frac{1+{\rm e}^{-\beta\hbar\omega}}
{\(1-{\rm e}^{-2\beta\hbar\omega}\)\(1-{\rm e}^{-\beta\hbar\omega}\)}.
\end{eqnarray}
The square root of this function, $\langle L_z^2\rangle^{1/2}/\hbar$,
shown in Fig.(\ref{C6LzRMS}), is an almost straight line 
as a function of $T/\hbar \omega$, with a small offset 
close to zero. 
Unity is reached at $T=0.76 ~ \hbar \omega$.
This means that the mode will contribute with angular momenta
quanta even for temperatures below the value corresponding 
to the vibrational quantum energies.
For the calculated lowest frequency of C$_6$ the value
$T=0.76 ~ \hbar \omega$ corresponds to a temperature of 
112 K, and for the highest frequency to 531 K. 
Both of these are fairly modest temperatures for carbon 
chemistry.

\section*{Summary}

Eq.\ref{Lzfinal} is the main results of this work. 
The numerical prescription derived should be applicable 
to any linear molecule with known vibrational frequencies. 
If the total vibrational level density specified with respect 
to both energy and angular momentum is needed, this is 
calculated by a simple convolution of the level densities 
for the modes that carry angular momentum with those that 
do not.

\section{Acknowledgements}
PF acknowledges a Postdoctoral grant from the Research 
Foundation – Flanders (FWO).

%\bibliography{Vib-L2ref}
%\bibliographystyle{apsrev}

\end{document}